\documentclass{elsart4}

\usepackage{graphicx}
\usepackage{amssymb}

\begin{document}
\begin{frontmatter}

\title{Precision neutron interferometric measurements of the n-p, n-d, and n-$^3$He zero-energy coherent neutron scattering
amplitudes}

\author[NCSU]{P.\,R. Huffman\corauthref{1}},
\author[NIST]{M. Arif},
\author[UNCW]{T.\,C. Black},
\author[NIST]{D.\,L. Jacobson},
\author[MISSOURI]{K. Schoen},
\author[IUCF]{W.\,M. Snow}, and
\author[MISSOURI,NIST]{S.\,A. Werner}

\address[NCSU]{North Carolina State University, Raleigh, NC 27695, USA}
\address[NIST]{National Institute of Standards and Technology, Gaithersburg, MD 20899, USA}
\address[UNCW]{University of North Carolina at Wilmington, Wilmington, NC 28403, USA}
\address[MISSOURI]{University of Missouri-Columbia, Columbia, MO  65211, USA}
\address[IUCF]{Indiana University/IUCF, Bloomington, IN 47408, USA}

\mbox{}\newline \small Paper presented as part of the Festschrift honouring Samuel A. Werner. \normalsize 

\corauth[1]{Corresponding Author: North Carolina State University, Department of Physics,
	    Campus Box 8202, Raleigh, NC 27695, USA.  Phone: (919) 515-3314
	    Fax: (919) 515-6238, Email: paul\_huffman@ncsu.edu}

\begin{abstract}
We have performed high precision measurements of the zero-energy neutron scattering amplitudes of gas phase molecular hydrogen,
deuterium, and $^{3}$He using neutron interferometry.  We find $b_{\mathit{np}}=(-3.7384 \pm 0.0020)$~fm\cite{Schoen03},
$b_{\mathit{nd}}=(6.6649 \pm 0.0040)$~fm\cite{Black03,Schoen03}, and $b_{n^{3}\textrm{He}} = (5.8572 \pm
0.0072)$~fm\cite{Huffman04}.  When combined with the previous world data, properly corrected for small multiple scattering,
radiative corrections, and local field effects from the theory of neutron optics and combined by the prescriptions of the Particle
Data Group, the zero-energy scattering amplitudes are: $b_{\mathit{np}}=(-3.7389 \pm 0.0010)$~fm, $b_{\mathit{nd}}=(6.6683 \pm
0.0030)$~fm, and $b_{n^{3}\textrm{He}} = (5.853 \pm .007)$~fm.  The precision of these measurements is now high enough to
severely constrain NN few-body models.  The n-d and n-$^{3}$He coherent neutron scattering amplitudes are both now in disagreement
with the best current theories.  The new values can be used as input for precision calculations of few body processes.  This
precision data is sensitive to small effects such as nuclear three-body forces, charge-symmetry breaking in the strong
interaction, and residual electromagnetic effects not yet fully included in current models.
\end{abstract}


\end{frontmatter}

\section{Introduction}

The last decade has seen a revolution in the accuracy with which low energy phenomena in nuclear few body systems can be
calculated.  Insight into certain features of few-nucleon systems has come both from greatly-improved calculations using potential
models based on the measured nucleon-nucleon (NN) interaction\cite{Carlson98} and also from the development of effective field
theory (EFT) approaches based on the chiral symmetry of QCD\cite{Bedaque98,Hammer99,Bedaque02}.  Such theories have been used to
develop a physical understanding rooted ultimately in QCD for the relative sizes of many quantities in nuclear physics, such as
nuclear N-body forces~\cite{Friar01} and in particular the nuclear 3-body force (3N), which is now investigated intensively.
Although it is well understood that 3N forces must exist with a weaker strength and shorter range than the NN force, little else
is known.

EFT has been used to solve the two and three nucleon problems with short-range interactions\cite{Bedaque02,Beane02}.  For the
two-body system, EFT is equivalent to effective range theory and reproduces its well-known results for NN
forces~\cite{Kolck98,Kaplan98,Gegelia98}.  The chiral EFT expansion does not require the introduction of an operator corresponding
to a 3N force until next-to-next-to leading order in the expansion, and at this order it requires only two low energy
constants~\cite{Epelbaum01,Epelbaum02}, which are taken to be the triton binding energy and the zero energy doublet n-d scattering
amplitude.  There have also been significant advances in other approaches to the computation of the properties of few-body nuclei
with modern potentials such as the AV18 potential~\cite{Wiringa95,Pieper01}, which includes electromagnetic effects and terms to
account for charge-independence breaking and charge symmetry breaking of the strong interaction.  These calculations accurately
reproduce the well-measured energy levels of few body bound states only with the phonomological inclusion of a nuclear 3-body
force, so it is clear that more information on this force is needed for further progress.

Precision measurements of low-energy strong interaction properties, such as the zero-energy scattering amplitudes and
electromagnetic properties of small $A$ nuclei, are therefore becoming more important for low energy, strong interaction physics
both as precise data that can be used to fix parameters in the EFT expansion and also as new targets for theoretical prediction.
In this article, we briefly summarize a series of precision measurements of the zero-energy coherent scattering amplitudes in the
two, three, and four-body systems using neutron interferometric techniques\cite{Schoen03,Black03,Huffman04}.  We summarize these
zero-energy coherent scattering amplitude results and discuss possibilities for future measurements.

\section{Neutron Optics Theory}
The zero-energy coherent scattering amplitude is the linear combination of scattering amplitudes that gives rise to the optical
potential of a neutron in a medium\cite{Rauch00}.  The zero-energy bound scattering amplitude, $b$, is related to the free
scattering amplitude $a$ by
\begin{equation}
    \label{eq:relofbtoa}
    b = \frac{m+M}{M}a.
\end{equation}
Here, $m$ is the mass of the neutron and $M$ is the mass of the atom.  For hydrogen, $a$ is the linear combination of
the singlet and the triplet scattering amplitudes given by,
\begin{equation}
    \label{eq:compnp}
    a_{\mathit{np}} = (1/4)~^{1}a_{\mathit{np}} + 
			(3/4)~^{3}a_{\mathit{np}},
\end{equation}
for deuterium it is the linear combination of the doublet and quartet scattering amplitudes,
\begin{equation}
    \label{eq:compnd}
    a_{\mathit{nd}} = (1/3)~^{2}a_{\mathit{nd}} + 
			(2/3)~^{4}a_{\mathit{nd}},
\end{equation}
and for $^{3}$He, it is the linear combination of the singlet and triplet scattering amplitudes,
\begin{equation}
    \label{eq:cohs1}
    a_{n^{3}\textrm{He}} = (1/4)~^{1}a_{n^{3}\textrm{He}} + 
				(3/4)~^{3}a_{n^{3}\textrm{He}}.
\end{equation}
The bound zero-energy scattering amplitude is one particular linear combination of the triplet and singlet (or doublet and
quartet) scattering amplitudes.  Knowledge of some other combination allows one to independently extract the individual bound (or
free) scattering amplitudes for each state.

The phase shift measured in neutron interferometry is proportional to the real part of the $S$-wave coherent scattering amplitude
in the medium,
\begin{equation}
    \phi=k(1-n)D=-\lambda NDb.
\end{equation}
where $k$ is the incident wave vector, $n$ is the index of refraction, $N$ is the number density, $D$ is the thickness of the
sample, and $\lambda$ is the neutron wavelength.  Thus to experimentally measure $b$, the neutron optical phase shift $\phi$, the
atom density, the sample thickness, and the neutron wavelength must each be measured to high precision.

\section{Experimental Procedure}
\begin{figure}[t]
    \begin{center}
    \includegraphics{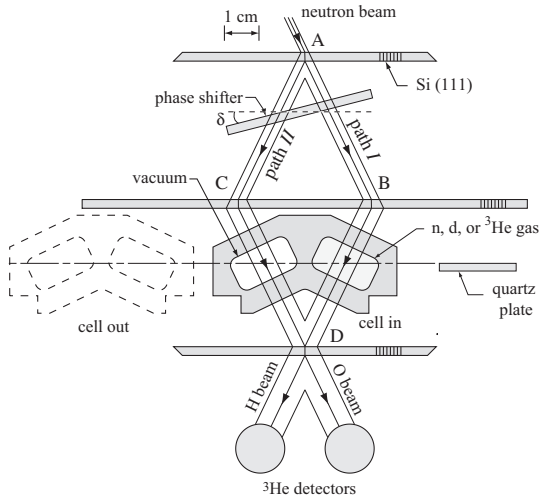}
    \end{center}
    \caption{A schematic view of the experimental setup as the neutron beam passes through the perfect crystal silicon
    interferometer.  Parameters associated with the neutron optics are discussed in the text.}
    \label{fig:Interferometer}
\end{figure}

Scattering amplitude measurements were performed at the National Institute of Standards and Technology (NIST) Center for Neutron
Research (NCNR) Interferometer and Optics Facility\cite{Ari94}.  A cold monochromatic neutron beam ($E = 11.1$~meV) enters the
perfect silicon crystal neutron interferometer and is coherently divided via Bragg diffraction into two beams that travel along
paths $I$ and $II$ as shown schematically in Fig.~\ref{fig:Interferometer}.  These beams are again diffracted and then coherently
recombined to form the interference pattern.  A detailed description of the facility, experimental arrangement, and procedures for
the determination of zero-energy neutron scattering amplitudes can be found in Ref.~\cite{Schoen03}.

A secondary sampling method is used to measure the phase shift $\phi$ due to the gas sample.  This is accomplished by positioning
a rotatable quartz phase shifter across the two beams as shown in Fig.~\ref{fig:Interferometer}.  The intensities of the beams
that arrive at the two $^3$He detectors are a function of the phase shifter angle $\delta$ and are given by
\begin{eqnarray}
    \label{eq:intensity}
    I_{O}(\delta) &=& A_{O} + B \cos(Cf(\delta)+ \phi_{gas} + \phi_{cell}) \\ \nonumber
    I_{H}(\delta) &=& A_{H} + B \cos(Cf(\delta)+ \phi_{gas} + \phi_{cell} + \pi).
\end{eqnarray}
The values of $A_{O}$, $A_{H}$, $B$, and $C$ are extracted from fits to the data.  The function $f(\delta)$ depends on the Bragg
angle $\theta_B$ and is a measure of the neutron optical path length difference between the beams induced by the phase shifter and
is given by
\begin{equation}
    \label{eq:pathdiff}
    f(\delta) = \frac{\sin(\theta_B)\sin(\delta - \delta_0)}{\cos^2(\theta_B) - \sin^2(\delta - \delta_0)}.  
\end{equation}

\begin{figure}[t]
    \begin{center}
    \includegraphics[width=0.45\textwidth]{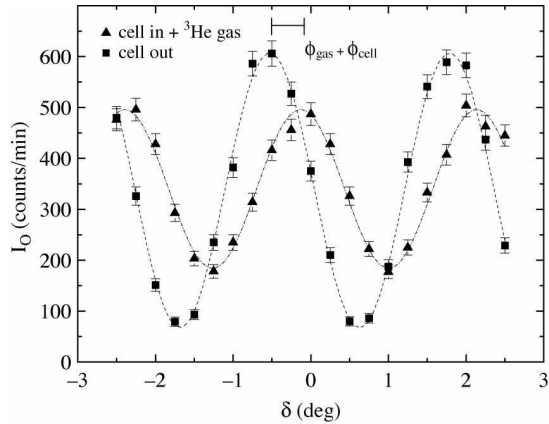}
    \end{center}
    \caption{A typical pair of interferograms with $^{3}$He present in the cell.  The oscillations arise from
    the change in path lengths created as the phase shifter is rotated (see Fig.~\ref{fig:Interferometer}).
    Data are shown for both the cell in and out of the interferometer.}
    \label{fig:interferogram}
\end{figure}

The hydrogen, deuterium, or $^{3}$He gases are housed in a cell specifically designed to minimize the phase shift $\phi_{cell}$
due to the aluminum walls of the cell (see Fig.~\ref{fig:Interferometer}).  The phase shifts arising from the presence of the gas
and cell ($\phi_{gas}$ and $\phi_{cell}$) were determined by collecting $\approx 10^{3}$ interferogram pairs with the cell
positioned both within the interferometer and removed from the beam paths.  The phase difference between cell-in/cell-out sets of
interferograms is extracted for each pair, with a typical set shown in Fig.~\ref {fig:interferogram}. The phase shift from the 
cell, $\phi_{cell}$, was determined using an evacuated cell.

The atom density was determined using the measured purity of the gas and the ideal gas law with virial coefficient corrections up
to the third pressure coefficient.  The absolute temperature was continuously monitored using two calibrated $100~\Omega$ platinum
thermometers that have an absolute accuracy of 0.023~\% at 300~K\@.  The pressure was continuously monitored using a calibrated
silicon pressure transducer capable of measuring the absolute pressure to better than 0.01~\%.  The wavelength of neutrons
traversing the interferometer was measured using a pyrolytic graphite (PG~002) crystal.  This analyzer crystal, calibrated
separately against a Si crystal with a precisely-known lattice constant, was placed in the H-beam of the interferometer and
rotated so  that both the symmetric and
anti-symmetric Bragg reflections were determined.  In a separate test, the stability of the wavelength over the measurement time
was shown to be
$0.001~\%$.  More details on the measurement techniques and systematic uncertainties involved
in the determination of the neutron wavelength, atom density, temperature, and cell thickness can be found in
refs.~\cite{Schoen03} and \cite{Huffman04}.

The value of the bound zero-energy scattering amplitude was calculated for each data set on a run-by-run basis.  These values were
combined using a weighted average to obtain $b$ for each gas species.  Our reported results are $b_{\mathit{np}}=(-3.7384 \pm
0.0020)$~fm\cite{Schoen03}, $b_{\mathit{nd}}=(6.6649 \pm 0.0040)$~fm\cite{Black03,Schoen03}, and $b_{n^{3}\textrm{He}} = (5.8572
\pm 0.0072)$~fm\cite{Huffman04}.

\section{Results and Discussion}

Since these measurements of the n-p, n-d, and n-$^{3}$He zero-energy scattering amplitudes were performed in an identical manner
using the same apparatus (cell, neutron wavelength analyzer, and pressure and temperature monitors), one can take the ratio of the
measurement values to obtain results which are even less sensitive to any potential remaining systematic errors.  Using $b_{np}$
as the reference, we obtain the ratios
\begin{eqnarray*}
    b_{n^{3}\textrm{He}}/b_{np} &=& (-1.5668 \pm 0.0021) \\
    b_{nd}/b_{np}               &=& (-1.7828 \pm 0.0014). 
\end{eqnarray*}
Although these ratios possesses slightly larger statistical uncertainties, it will be independent of any unknown systematic
uncertainty to first order and can provide a more robust target for comparison to theories which attempt to calculate all of the
scattering lengths within a common theoretical framework.

\begin{figure}[t]
    \begin{center}
    \includegraphics[width=2.75in]{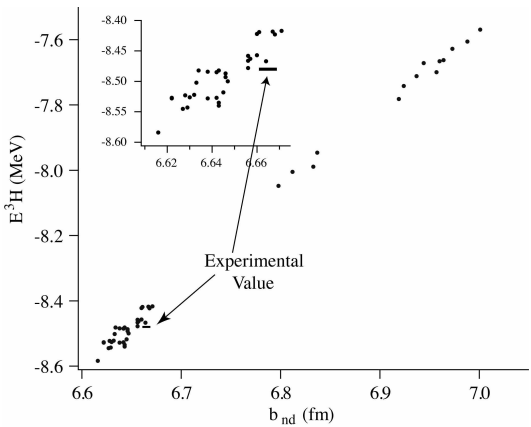}
    \end{center}
    \caption{Theoretically determined values for the n-d scattering amplitude $b_{\mathit{nd}}$ and the triton binding energy
    using different NN and 3N forces\cite{Wit03}.  The experimental value consists of our measurement of $b_{\mathit{nd}}$ and the
    very precisely known value of the triton binding energy, $(8.481855 \pm 0.000013)$~MeV\cite{Wap83}.}
    \label{fig:theory1}
\end{figure}

Recently, theoretical predictions of the n-d zero-energy scattering amplitude were performed using the high precision NN forces CD
Bonn 2000, AV18, Nijm I, II and 93 in combination with 3N force models\cite{Wit03}.  The results of these calculations are
summarized in Fig.~\ref{fig:theory1} alongside the experimental data.  For NN forces alone with and without electromagnetic
interactions, they recovered the approximate correlation between the triton binding energy and $^{2}a_{\mathit{nd}}$ known as the
Phillips line.  The approximate correlation between these observables is now understood to be a generic feature for systems such
as the deuteron and other low A nuclei whose size is significantly larger than the ~1 fm scale of the NN interaction range.
However it is understood that this correlation is only an approximation and should not be obeyed exactly: for example 3N forces
will introduce corrections.  Although the addition of 3N forces of a wide variety of types does shift the values closer to the
observed $^{2}a_{\mathit{nd}}$, none of these calculations agrees with the new high precision measurements.  The authors note that
the $^{2}a_{\mathit{nd}}$ zero-energy scattering amplitude must be considered as an independent low energy observable for future
calculations in few body systems.

\begin{figure}[t]
    \begin{center}
    \includegraphics[width=2.75in]{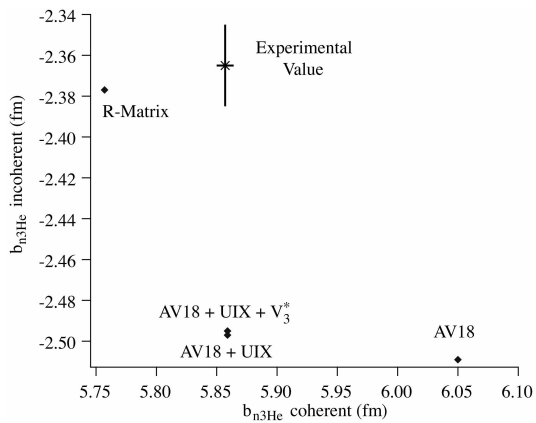}
    \end{center}
    \caption{Comparisons of the n-$^{3}$He coherent and incoherent scattering amplitudes as determined from our measurement and
    the measuremetn of Zimmer \textit{et al.}\cite{Zim02} to theoretical calculations of these same parameters using either an
    R-matrix formalism or the AV18 NN potential with two types of 3N forces\cite{Hof03}.}
    \label{fig:theory2}
\end{figure}

A second group has recently published new calculations of the spin-dependent $n-^3$He scattering amplitudes using the Resonating
Group Method and a variety of modern NN and 3N potentials\cite{Hof03}.  Comparisons of the experimental free nuclear singlet and
triplet scattering amplitudes as determined from our measurement and the n-$^{3}$He incoherent scattering amplitude from Zimmer
\textit{et al.}\cite{Zim02} with theoretical calculations of these parameters were performed using an R-matrix formalism and the
AV18 NN potential with two types of 3N forces and shown in Fig.~\ref{fig:theory2}.  It is clear that the theoretical predictions
from even these impressive, state-of-the-art calculations using resonating group techniques still lie outside the range of
experimental uncertainties for the $^3$He binding energy and the singlet and triplet n-$^{3}$He scattering amplitudes.

The precision of these measurements is now high enough to severely constrain these few-body models.  Both the n-d and n-$^{3}$He
coherent neutron scattering amplitudes are in disagreement with the best current theories.  Only models which correctly take into
account nuclear three-body forces, charge-symmetry breaking, and residual electromagnetic effects have the possibility to
successfully confront the data.

We can combine these measurements with the previous world's data to obtain new values for the coherent scattering amplitudes .  We
note that some of the past measurements of high precision using a gravity reflectometer performed in Garching a few decades
ago\cite{Koe91} must be corrected for local field effects\cite{Sears85}, which are significant for the coherent neutron scatting
amplitude in hydrogen $b_{\mathit{np}}$ as measured in the gravity refractometer but are negligible in neutron interferometry.
This correction shifts the value from $b_{\mathit{np}} = (-3.7409\pm0.0011)$~fm\cite{Koe91} to $b_{\mathit{np}} =
(-3.7390\pm0.0011)$~fm\cite{Sears85}, an approximately two sigma effect.  In addition there is another two sigma correction to the
neutron interferometry measurement in hydrogen gas due to multiple scattering and correlation corrections to the neutron index of
refraction which are negligible for the gravity refractometer measurements.  These corrections have already been applied to obtain
the value of $b_{\mathit{np}}=(-3.7384 \pm 0.0020)$~fm\cite{Schoen03} quoted above.  After these corrections, both independent
measurements are in excellent agreement.  We can therefore present an improved value of the coherent neutron scattering length of
hydrogen of $b_{\mathit{np}}=(-3.7389 \pm 0.0010)$~fm.  These corrections turn out to be negligible for deuterium.

\section{Future Measurements}
As mentioned above, the measured value of $b_{\mathit{np}}$ was corrected for multiple scattering and correlation corrections to
the neutron index of refraction.  A more precise interferometric measurement in comparison with the gravity refractometer value
would be able to isolate this term experimentally.  Such a measurement would constitute the first experimental observation of
corrections to the neutron index of refraction due to virtual excitations and multiple internal scattering within a molecule.
Such effects were predicted by Nowak 20 years ago\cite{Now82b}.  The calculation of Nowak, which was done in the long wavelength
limit $kR_{0} \rightarrow 0$ ($R_{0}$ = bond length of H$_{2}$ = 0.74611 nm, for D$_{2}$ $R_{0}$ = 0.74164 nm \cite{CRC}), must be
extended to the conditions of the experiment (which correspond to $kR_{0}$=1.73) to make a precise prediction.  It would be
possible to measure $b_{\mathit{np}}$ using H$_{2}$ gas at least a factor of two more precisely, which may be accurate enough to
isolate this neutron optics effect experimentally for the first time.
 
Our measurement of $b_{\mathit{nd}}$ will hopefully soon be complemented by a measurement in progress at PSI of the incoherent n-d
scattering amplitude using pseudomagnetic precession in a polarized deuterium target\cite{Bra04}, which is also an interferometric
technique that operates in neutron spin space as opposed to real space.  This measurement will allow one to separate the two spin
channels and determine the interesting doublet amplitude, $^{2}a_{\mathit{nd}}$, with high precision.  We have argued that the
quartet amplitude should be independent of 3N forces and the theoretical value should be reasonably robust\cite{Fri00} and were
thus able to extract a value of $^{2}a_{\mathit{nd}}=(0.645\pm0.003$(expt)~$\pm~0.007$(theory))~fm\cite{Huffman04}, however an
experimental determination is badly needed.  We expect that these new measurements will improve $^{2}a_{\mathit{nd}}$ by an order
of magnitude to $\approx 10^{-3}$.  The doublet amplitude is very important from a theoretical point of view: in the EFT approach
it fixes one of the two low energy constants in the expansion to next-to-next-to leading order (NNLO) and its determination will
allow more precise predictions to be made for other few body systems.

It is also interesting to consider how the experimental determinations of the n-$^{3}$He scattering amplitudes can be further
improved.  Our measurement of the zero-energy scattering amplitude is dominated by statistical uncertainties in the measurement of
the phase shifts and consequently there is still room for improvement.  In the case of the incoherent scattering amplitude
($b_{i}$) determination from pseudomagnetic precession performed by Zimmer \textit{et al.}\cite{Zim02}, the accuracy is
unfortunately limited by the poor experimental knowledge of the relative contributions of singlet and triplet channels to the
n-$^{3}$He absorption cross section.  A better measurement of this ratio, currently known to $\approx 1$~\%\cite{Pas66,Bor82},
could be immediately combined with the Zimmer \textit{et al.} measurement to improve the accuracy of $b_{i}$ by as much as a
factor of three.  In addition, a new measurement currently being planned at NIST to directly measure the spin-dependent n-$^{3}$He
scattering amplitudes is independent of this ratio \cite{Fred}.  Dramatically reduced uncertainties for $^{1}a_{n^{3}\textrm{He}}$
and $^{3}a_{n^{3}\textrm{He}}$ in n-$^{3}$He are therefore possible.

Measurements in other light nuclei are also quite feasible.  Two targets that we are presently exploring are $^{4}$He and tritium.
From a theoretical point of view at present, a  $^{4}$He measurement is not that interesting because the framework for solving
five-body problems is not yet in place.  The tritium measurement on the other hand is quite interesting theoretically because the
small inelastic effects make it much easier to calculate than n-$^{3}$He, but it is experimentally difficult because of the
radioactive nature of the target.  We are investigating the possibility of performing a measurement of $b_{\mathit{nt}}$ using the
same techniques used in the current measurements.  We expect that the limiting factor in these measurements will be how well we
can determine the isotopic composition of the radioactive tritium gas.
 
Perhaps the single most interesting low energy NN scattering amplitude to measure is the neutron-neutron scattering amplitude
$b_{\mathit{nn}}$.  A value for $b_{\mathit{nn}}$ is the last remaining obstacle to theoretically calculating the n-A
scattering amplitudes ($\mbox{A} > 3$) from first principles.  At present, no direct measurements of $b_{\mathit{nn}}$ exist.  An
experiment to determine $b_{\mathit{nn}}$ by viewing a high-density neutron gas near the core of a reactor and measuring a
quadratic dependence of the neutron fluence on source power is currently being designed\cite{Furman02}.  A second experiment to
let the neutrons in an extracted beam scatter from each other has also been considered~\cite{Pokotolovskii93}.  An EFT analysis to
extract $b_{\mathit{nn}}$ from the $\pi^{-}d \to nn\gamma$ reaction has also recently been performed~\cite{Gardestig05}.  Any
direct measurement would need to be performed to an accuracy of a few percent to be physically interesting.

One additional measurement using neutron interferometry is presently underway at NIST, a determination of the neutron's mean
square charge radius by measuring the neutron-electron scattering amplitude, $b_{\mathit{ne}}$\cite{Wie05}.  This experiment is
expected to lead to a five-fold improvement in the precision of $b_{\mathit{ne}}$.  This precision may be sensitive to radiative
corrections calculable in a recent EFT approach\cite{Pineda04}.

Theoretical advances over the last decade have made few-nucleon systems into a quantitative testing ground for low energy QCD. Low
energy neutrons can be used to perform high-precision measurements of scattering amplitudes.  We look forward to new
high-precision measurements in this field in the next few years.

\section{Acknowledgements}
We acknowledge the support of the National Institute of Standards and Technology, U.S. Department of Commerce, in providing the
neutron research facilities used in this work.  This work was supported in part by the U.S.\ Department of Energy under Grant No.\
DE-FG02-97ER41042 and the National Science Foundation under Grants No.\ PHY-9603559 at the University of Missouri, No.\
PHY-9602872 at Indiana University, and No.\ PHY-0245679 at the University of North Carolina at Wilmington.

\end{document}